\documentstyle[aps,twocolumn,prl,tighten,epsfig]{revtex} 
\input epsf
\def\plotone#1{\centering \leavevmode
\epsfxsize= 0.95\columnwidth \epsfbox{#1}}

\def\TeV{\,{\rm TeV}}

\def\Mpc{\,{\rm Mpc}}

\def\eV{{\,\rm eV}}

\def\cmm2{{\,\rm cm^{-2}}}
\def\cm2{{\,{\rm cm}^2}}
\def\cmm3{{\,{\rm cm}^{-3}}}
\def\gcmm3{{\,{\rm g\,cm^{-3}}}}

\def\mpl{{m_{\rm Pl}}}

\def\la{\mathrel{\mathpalette\fun <}}
\def\ga{\mathrel{\mathpalette\fun >}}
\def\fun#1#2{\lower3.6pt\vbox{\baselineskip0pt\lineskip.9pt
  \ialign{$\mathsurround=0pt#1\hfil##\hfil$\crcr#2\crcr\sim\crcr}}}
\def\lcdm{$\Lambda$CDM~}

\def\ie{{\it i.e.}}

\begin{document}
\twocolumn[\hsize\textwidth\columnwidth\hsize\csname @twocolumnfalse\endcsname

\title{Solving the Coincidence Problem: Tracking Oscillating Energy} 

\author{Scott Dodelson$^{1,2}$,
 Manoj Kaplinghat$^{2}$, and Ewan Stewart$^{1,3}$
}

\address{$^1$NASA/Fermilab Astrophysics Center
Fermi National Accelerator Laboratory, Batavia, IL~~60510-0500}
\address{$^2$Department of Astronomy \& Astrophysics
Enrico Fermi Institute, The University of Chicago, 
Chicago, IL~~60637-1433}
\address{$^3$Department of Physics, KAIST, Taejon 305-701,
South Korea}

\date{\today}
\maketitle

\begin{abstract}
Recent cosmological observations strongly suggest that the universe
is dominated by an unknown form of energy with negative pressure.
Why is this dark energy density of order the critical density today? 
We propose that the dark energy has 
periodically dominated in the past so that its preponderance 
today is natural. We illustrate this paradigm with a model potential 
and show that its predictions are consistent with all observations.
\end{abstract}
]

{\parindent0pt\it Introduction.}
A variety of evidence accumulated over the last several years 
points to the existence of an unknown, unclumped form of energy in the Universe.
First was an apparent concordance \cite{Concordance} of different measurements: 
the age of the Universe; the Hubble constant; the baryon fraction in clusters;
and the shape of the galactic power spectrum.
Second came the stunning observations \cite{SN} of tens of distant Type Ia Supernovae,
which found a distance-redshift relation in accord with a cosmological constant,
but in strong disagreement with a matter dominated Universe.
Finally, this past year has seen analyses \cite{cmb} of the experiments 
measuring anisotropies in the CMB.
Taken together, the CMB experiments plot out a rough shape for the power spectrum,
one that is in accord with a flat Universe, but in disagreement with an open Universe.
If we believe the estimates of matter density coming from observations of clusters \cite{cluster},
the only way to get a flat Universe, and hence account for the CMB measurements,
is to have an unclumped form of energy density pervading the Universe.

Perhaps the simplest explanation of these data is that the unclumped form of energy density
corresponds to a positive cosmological constant\cite{coscon}.
A non-zero but tiny constant vacuum energy density (cosmological constant) could
conceivably be explained by some unknown string theory symmetry
(that sets the vacuum energy density to zero) being broken by a small amount.
However, to explain in this way a constant vacuum energy density of
$2 \times 10^{-59} \TeV^4$, which is not only small but is also just the right value
that it is just beginning to dominate the energy density of the Universe {\em now\/},
would require an unbelievable coincidence.
A different possibility is to 
give up the dream of finding a mechanism which would set the vacuum
energy density to exactly zero and resort to believing that anthropic considerations
select amongst $\gtrsim 10^{100}$ string vacua to find one with a vacuum energy
density sufficiently fine-tuned for life.
Although this anthropic selection mechanism is logically consistent and even predicts
a small but observable cosmological constant, one might think that nature would have
found a more efficient mechanism to obtain a sufficiently small cosmological constant
than such extreme brute force application of anthropic selection.

An alternative is to assume that the true vacuum energy density is zero,
and to work with the idea that the unknown, unclumped energy is due to a scalar field
$\phi$ which has not yet reached its ground state.
This idea, which is called dynamical lambda or quintessence, has received much
attention \cite{quint} over the last several years.
However, two problems still remain.
First, the field's mass has to be extremely small, less than or of order the Hubble
constant today $\sim 10^{-33} \eV$, to ensure that it is still rolling to its vacuum
configuration.
This is in general difficult because scalar fields tend to acquire masses greater than
or of order the scale of supersymmetry breaking suppressed by at most the Planck scale:
$ m \gtrsim F/\mpl \gtrsim {\rm TeV}^2 / \mpl \sim 10^{-3} \eV$.
Although difficult, this could be achieved using pseudo-Nambu-Goldstone bosons
\cite{Goldstone}.
Another more speculative way to achieve this would be to use the hypothetical symmetry
(perhaps some sort of hidden supersymmetry) that ensures that the true vacuum energy
density is zero to also protect the flat directions in scalar field space that would
correspond to the very light scalar fields necessary for quintessence.
The second, and perhaps even more serious problem is that almost all of these models
require that we live in a special epoch today, when the quintessence is just starting
to dominate the energy density of the Universe, and furthermore this specialness cannot
even be justified by use of anthropic arguments.

In recent years a lot of progress has been made in understanding the
behavior of quintessence fields. A broad class of solutions, called tracker 
solutions \cite{steinhardt}, has been discovered in which the final value 
of the quintessence energy density is insensitive to the initial conditions.
For example, potentials like $V = V_0 \phi^{-n}$ or $V = V_0 \exp(1/\phi)$ 
can, for suitable choices of $V_0$, catch up with the critical density late 
in the evolution of the Universe for a wide range of initial conditions
and thus provide a natural setting for explaining the current acceleration 
of the Universe. However, the suitable choice of $V_0$ must be of the order 
of the critical energy density today, \ie, we are back to the 
problem of living at a special epoch today and not even being able to use 
anthropic arguments to justify this specialness. 

In a subset of these tracking models, which we call the exact tracker 
solutions \cite{tracker,joyce}, the scalar field energy density is always 
related to the ambient energy density in the Universe: if the 
dominant component in the Universe is radiation, then the tracking 
field's energy density also falls off as $a^{-4}$, where $a$ is the 
scale factor of the Universe. If the dominant component is matter, then 
the field's energy density scales as $a^{-3}$. This behavior arises from 
an exponential potential for $\phi$ (regardless of the value of $V_0$). 
Since the energy density in this field is always comparable to the 
background density, we are not living at a special epoch: any observer 
in the distant past or future would also see the tracking field's energy 
density. However, these tracking solutions run into two problems. First, 
if their energy density today truly is dominant, then it should also have 
been dominant at the time of Big Bang Nucleosynthesis (BBN). Constraints 
from observations of light element abundances preclude such an additional 
form of energy density at early times. Second, tracking models have the 
wrong equation of state at present since the tracking field behaves 
like matter, with zero pressure, instead of having the necessary negative 
pressure to accelerate the Universe.

In this {\em letter} we ask the question, what if the Universe has been
accelerating periodically in the past? Then the fact that the Universe is 
accelerating today would not be surprising. It would merely reflect
that the period is such that the Universe is accelerating today.
Of course, if it turned out that to achieve a presently accelerating
Universe the period had to be excessively fine-tuned, then this scenario 
would not be worth considering. However, note that the assumption that there
is nothing special about the present time itself argues for the robustness of 
such a scenario. If the Universe does accelerate periodically, then there
is no reason why it should not be accelerating today. If the Universe
does accelerate periodically, then it is, in fact,  reasonable to expect it 
to accelerate today. 

To judge the merits of this scenario in a concrete manner, we adopt an
ad-hoc potential. Though worked out for this specific potential, the 
predictions outlined here are the generic predictions of a periodically 
accelerating Universe. The model we adopt for study is a modification of 
the exponential potential (which leads to the exact tracker solution).
The modification to the potential is a sinusoidal modulation, which
induces the tracker field to oscillate about the ambient energy density.
We show that such a potential can satisfy the the BBN constraints, can produce 
the right equation of state today and leads to testable features in the 
CMB and matter power spectra.
We call this type of energy {\em Tracking, Oscillating Energy}, or TOE.

{\parindent0pt\it The potential and the field evolution.}
Consider a scalar field $\phi$ with potential
$V(\phi) = V_0 \exp( - \lambda \phi \sqrt{8\pi G})$.
It is well-known \cite{tracker} that such a potential leads to an attractor solution with
$\Omega_\phi \equiv \rho_\phi /(\rho_\phi + \rho_o) = n/\lambda^2$
where $\rho_o$ is the energy density in the other component of the Universe,
which is assumed to scale as $a^{-n}$. Thus, no matter what the initial
conditions are for $\phi$, it always evolves so that it tracks the
rest of the density in the Universe. 

\begin{figure}[thbp]
\plotone{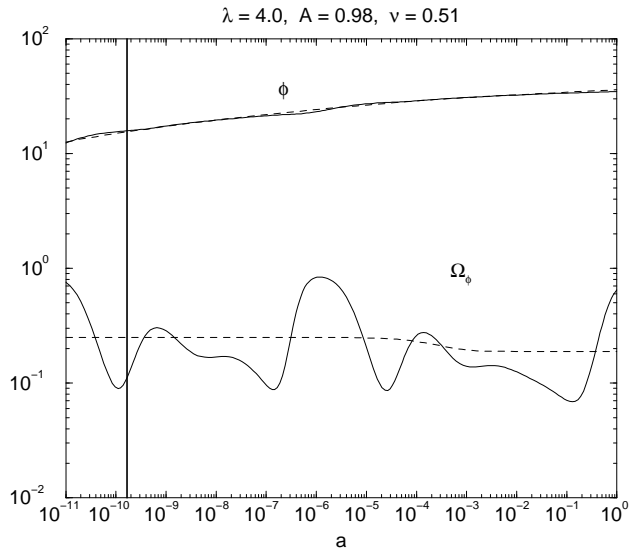}
\caption[caption]{The fraction of the critical density in $\phi$
for the potential in Eq.~(\ref{pot}). The dotted line shows the corresponding tracking
solution ($A=0$). The upper set of curves shows the evolution in 
$\phi$ for the TOE and the tracking models.}
\label{fig1}
\end{figure}

Now consider the potential
\begin{equation}\label{pot}
V(\phi) = V_0 \exp \left( -\lambda\phi \sqrt{8\pi G} \right)
\left[ 1 + A \sin(\nu\phi\sqrt{8\pi G}) \right].
\end{equation}
This potential serves to modulate the tracking behavior.
Figure \ref{fig1} shows the resultant evolution of $\phi$
and its energy density for a particular set of the parameters
$A,\nu$. (The normalization $V_0$ can be set to $G^{-2}$
by shifting the initial value of $\phi$.)
Also shown is the tracking solution for this
particular value of $\lambda$ without the modulation. 
As expected, the sinusoidal term in the potential leads to
oscillations about this tracking behavior. 
One can obtain analytic solutions for the dynamics of the potential
in Eq.~(1)
during radiation ($n=4$) or matter ($n=3$) domination in the
limit that $A$ is small by perturbing about the corresponding exact tracker
model which has $\phi\sqrt{8\pi G} = \frac{n}{\lambda} \ln a$. The sine in Eq.~(1) provides
a periodic forcing term with period $\ln a = \frac{2 \pi \lambda}{n
\nu}$,
while the natural period\cite{tracker} of the damped oscillations about the exact
tracker solution is $\ln a = 8 \pi \lambda / \sqrt{(6-n)[3(3n-2)\lambda^2-8n^2]}$
with decay $e$-life $\ln a = 4/(6-n)$.
Although the above results are strictly valid only for small $A$, they
account remarkably well for the behaviour shown in Figure~1. The forced
period corresponds to the longer period of $5.4$ units ($n=4$) and
somehwere between $5.4$ units and $7.1$ units ($n=3$), while the natural
period corresponds to the shorter period of $1.6$ units ($n=4,3$) of the
damped oscillations which are presumably excited by the non-linear
effects
that appear when $A$ is not small.

The energy density due to $\phi$ is relatively small at the time
of BBN and relatively large today for the parameter set in Figure~\ref{fig1}.
It is, of course, clear that in order to get the right behavior at BBN and 
today, one has to pick the ``correct'' parameter sets. This involves a bit of 
fine-tuning which, as we argue below, is quite reasonable and natural.
If one thinks of the parameter set as being randomly selected, then there 
is a finite probability that the Universe will be accelerating today and 
that the energy density of $\phi$ will be sub-dominant at BBN. What is this 
probability? If one selects $A$, $\nu$ and $\lambda$ randomly, the chance 
of getting a Universe like ours is of the order of 1 in a 100. The exact 
number (for this potential) depends on how stringently we define 
``a Universe like ours''. For example the tight constraints 
$0.4 < \Omega_\phi < 0.8$, $w_\phi < -0.5$, and 
$(\rho_\phi/\rho_0)_{\rm BBN} < 0.1$ give a probability of 1 in 
450, while the relaxed constraints $0.1 < \Omega_\phi < 0.9$ and 
$w_\phi < -0.25$ and $(\rho_\phi/\rho_0)_{\rm BBN} < 0.2$
give a probability of 1 in 26. It is also very important
to note that whatever the extent of fine-tuning, all of it is in 
{\em dimensionless} numbers. There are no energy scales in this scenario 
which are to be set by the present expansion rate of the Universe.

{\parindent0pt\it Power Spectra.}
To compare with CMB and large scale structure observations, we compute 
the power spectra of the perturbations in a TOE model.
Perturbations evolve differently in the presence of the scalar field 
energy density. For example, perturbations typically grow only when 
the Universe is matter dominated. Therefore, we expect a non-zero
$\Omega_\phi$ to lead directly to power suppression on the scales inside the 
horizon, with increased suppression for larger $\Omega_\phi$. 

\begin{figure}[thbp]
\plotone{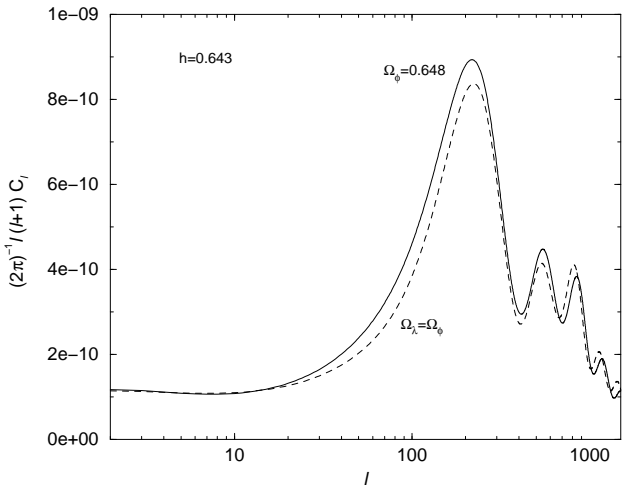}
\caption[caption]{The angular photon power spectrum from the TOE model of 
Figure \ref{fig1}. Also shown is a cosmological constant model with 
all other parameters equal.}
\label{cls}
\end{figure}

The prediction for the CMB angular power spectrum is plotted in Figure 
\ref{cls}. The primeval power spectrum is scale-invariant with adiabatic 
initial conditions. Also plotted for comparison is a model ($\Lambda$CDM) with
cosmological constant $\Omega_\Lambda=\Omega_\phi$ today and the rest of the 
cosmological parameters
also being the same. In further discussions we will contrast the results from
the TOE model against this \lcdm model. 
A noteworthy feature in Figure \ref{cls} is the increase in the heights of 
the first two peaks compared to that of the \lcdm model. This stems from
the fact that the gravitational potential decays 
more in the presence of the additional quintessence energy density. The decay of the
potential at and after recombination (the so-called Integrated Sachs-Wolfe
, or ISW, effect) leads\cite{HS} to enhanced power on scales $l \la 600$,
after which the potential becomes irrelevant. Note that the increase in the amplitude of both
the first and second peak cannot be mimicked by adding more baryons, which raise the odd peaks
but lower the even ones.

On smaller scales $(l\ga 600)$, the TOE model has smaller anisotropies. Here there are
two competing effects. First, the 
difference between the TOE and the \lcdm models (around recombination
when $\Lambda$ is insignificant) is the presence of the extra quintessence 
energy density, which leads to the expansion rate in the two models
being related as--
\begin{equation}
H_{\mathrm{TOE}}(a) = H_{\mathrm{\Lambda CDM}} (a)\; \times\; 
\left(1-\Omega_\phi(a)\right)^{-1/2}\;.
\label{expansion}
\end{equation}
Eq. \ref{expansion} implies that all the relevant scales at recombination 
(which occurs at $a_r\simeq 10^{-3}$) are smaller in the TOE model by a 
factor of about $\sqrt{1-\Omega_\phi(a_r)}$. In particular, the damping 
scale is smaller, which increases in the power on small
scales for the TOE model relative to the \lcdm model. The second effect is 
the large scale normalization of the two models \cite{footnote1}, 
and this second effect more than
compensates for the first. COBE normalization is sensitive to scales around 
$\ell=10$ for which the differences in the two models with regard to the 
late-ISW effect is important. In particular, since $\Lambda$ domination 
occurs very late, the ISW contribution around $\ell=10$ is much larger in
the TOE model. This in turn implies that the normalization of the
primeval power spectrum is smaller, a fact noticeable in the smaller amplitude
of the photon power spectrum for the TOE model at small scales (and also the matter
power spectrum, as we will soon see). 
One last effect that is worth pointing out concerns the difference in
the peak positions in the two models (though unlike the peak amplitudes, it
is probably not easily discerned). In particular, the TOE model has the acoustic 
features in its angular power spectrum shifted to smaller scales. This 
directly traces to the decrease in the angular diameter distance to the 
last scattering surface, for the TOE model. Of course, there is also 
the competing effect of the decrease in the size of the sound horizon at 
last scattering for the TOE model, which minimizes the effect.

\begin{figure}[thbp]
\plotone{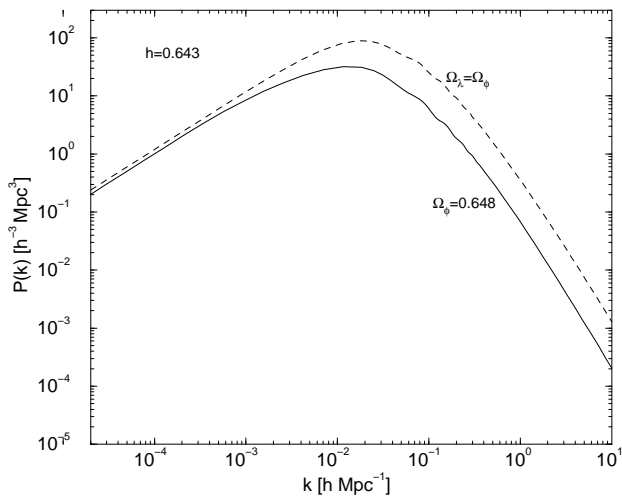}
\caption[caption]{The matter power spectrum from 
the TOE model of Figure \ref{fig1}. 
Also shown is a cosmological constant model with 
all other parameters equal. Power is significantly smaller in the TOE model.}
\label{ps}
\end{figure}

The prediction for the matter power spectrum is plotted in Figure \ref{ps}.
The difference in power at the largest scales is due to COBE normalization
and the difference in the super-horizon growth factor (which is sensitive 
to the equation of state of the cosmic fluid) for the perturbation. 
As one moves to smaller scales, which entered the horizon well before 
the present, the differences in the evolution of the matter perturbation
become more pronounced. The presence of the extra quintessence energy 
stunts the growth of perturbation once a mode enters the horizon. So, the 
earlier the mode enters the horizon, the larger the growth suppression 
relative to the \lcdm model. In other words, smaller modes are 
monotonically
more suppressed (something that may not be noticeable in the log plot) 
compared to the same modes in \lcdm model. It might also be surprising that
the $\phi$ domination around $a=10^{-6}$ does not cause a more appreciable
feature (\ie, suppression) in the power spectrum. The reason is that the
smallest scales in Figure \ref{ps} have just entered the horizon at the
time of $\phi$ domination ($a\sim 10^{-6}$).

The normalization on the small scales is generally quoted in terms of 
$\sigma_8$, the rms mass fluctuation within a $8\,h^{-1}\Mpc$ sphere. For the
parameters in Figure \ref{fig1}, the TOE model has $\sigma_8=0.4$. This
is several sigma smaller than the preferred value (see e.g. \cite{wang}) of $\sim 0.8$,
but could be rectified by a small blue-shift in the primordial spectrum
\cite{stewart}.

{\parindent0pt\it Conclusions.} We have constructed a model
wherein the energy density tracks the dominant component
in the Universe; satisfies the BBN constraints; and has
the proper equation of state today. Further, this model
makes definite predictions for large scale structure and for the
CMB.

Perhaps the greatest drawback of this class of models is the
arbitrariness of the potential.
In particular we know of no theory which predicts a potential
of the form given in Eq.~(\ref{pot}).
Nonetheless, we feel that the testable predictions of the model
and the aesthetic quality it preserves that we do not live in
a special epoch are of sufficient interest to warrant further
study.

\bigskip

We thank Limin Wang for helpful discussions.
The CMB spectra used in this work were generated by an
amended version of CMBFAST 
\cite{CMBCalculations}.
This work was supported by the DOE and the NASA grant NAG
5-7092 at Fermilab.
EDS acknowledges support by the KOSEF
Interdisciplinary Research Program grant 1999-2-111-002-5
and the Brain Korea 21 Project.

\newcommand\sapj[3]{ {\it Astrophys. J.} {\bf #1}, #2 (19#3) }
\newcommand\sprd[3]{ {\it Phys. Rev. D} {\bf #1}, #2 (19#3) }
\newcommand\sprl[3]{ {\it Phys. Rev. Letters} {\bf #1}, #2 (19#3) }
\newcommand\np[3]{ {\it Nucl.~Phys. B} {\bf #1}, #2 (19#3) }


\begin{thebibliography}{99}
\bibitem{footnote1} Note that the normalization affects the first two peaks 
also: the TOE model would have an even larger relative amplitude there if 
not for the large scale normalization.
\end{thebibliography}
\end{document}